# Conceptual Framework and Documentation Standards of Cystoscopic Media Content for Artificial Intelligence

Okyaz Eminaga (1,2) Timothy Jiyong Lee (1), Jessie Ge (1), Eugene Shkolyar (1), Mark Laurie (1), Jin Long (2), Lukas Graham Hockman (1), Joseph C. Liao (1,2)

1) Department of Urology, Stanford University School of Medicine, Stanford
2) Center for Artificial Intelligence and Medical Imaging, Stanford University School of Medicine, Stanford, CA

Statements and Declarations

The Authors declare no Competing Financial or Non-Financial Interest.

Corresponding authors

Joseph C. Liao, MD

Department of Urology, Stanford University School of Medicine, Stanford, CA, USA

Email: jliao@stanford.edu

Okyaz Eminaga, MD/PhD

Department of Urology, Stanford University School of Medicine, Stanford, CA, USA

Email: okyaz.eminaga@stanford.edu

Department of Urology

Stanford University School of Medicine,

453 Quarry Road, Mail Code 5656

Palo Alto, California, 94304, USA

# Abstract

**Background:** The clinical documentation of cystoscopy includes visual and textual materials. However, the secondary use of visual cystoscopic data for educational and research purposes remains limited due to inefficient data management in routine clinical practice.

**Methods:** A conceptual framework was designed to document cystoscopy in a standardized manner with three major sections: data management, annotation management, and utilization management. A Swiss-cheese model was proposed for quality control and root cause analyses. We defined the infrastructure required to implement the framework with respect to FAIR (findable, accessible, interoperable, re-usable) principles. We applied two scenarios exemplifying data sharing for research and educational projects to ensure the compliance with FAIR principles.

**Results:** The framework was successfully implemented while following FAIR principles. The cystoscopy atlas produced from the framework could be presented in an educational web portal; a total of 68 full-length qualitative videos and corresponding annotation data were sharable for artificial intelligence projects covering frame classification and segmentation problems at case, lesion and frame levels.

**Conclusion:** Our study shows that the proposed framework facilitates the storage of the visual documentation in a standardized manner and enables FAIR data for education and artificial intelligence research.

## Introduction

Cystoscopy is a common endoscopic procedure that visualizes the interior of the bladder and urethra. Cystoscopy is utilized in the diagnosis and treatment of many lower urinary tract diseases including benign prostatic hyperplasia, urethral stricture, and bladder cancer, and is often performed at regular intervals for disease surveillance. The accurate documentation of cystoscopy findings is essential to provide a longitudinal view of disease, but this has primarily relied on the text-based documentation of visual information.

Modern endoscopy equipment is capable of recording images and videos of the procedure to provide visual supplementation to text-based documentation of endoscopic findings, but this is not routinely stored in electronic health records for a variety of reasons like the lack of readiness or the inadequate implementation strategy [1,2]. In contrast to radiological imaging such as plain X-ray, CT, MRI, and ultrasound, endoscopic imaging lacks a standardized framework to store and catalog images and videos in the medical record. Inefficient management and inconsistent quality of cystoscopic data result in limited secondary use for education and research. Timely extraction of high-quality data from cystoscopic imaging requires the synergic effort of a multidisciplinary team[3], and raises the need for a standardized conceptual framework to facilitate this in an efficient manner [4,5].

With the rise of artificial intelligence (AI) and online education platforms for urology, cystoscopic media contents are gaining momentum due to their relevance for AI and education. Moreover, recent studies demonstrate encouraging results in the utility of AI as an adjunct tool for cystoscopy [6-10]. Details on data management remain unclear in much of the current literature, and as consequence the data sharing potential of these studies is limited. Accordingly, standardized documentation is crucial for efficient data management and exchange [11]. In the era of artificial intelligence, visual data has further become an increasingly crucial asset to develop machine learning (ML) algorithms for clinical decision making [12,13].

With explosive production and sharing of scientific data over the last decade, FAIR principles haven been proposed to improve the management of scientific data; FAIR stands for Findability, Accessibility, Interoperability, and Reusability and one of its aims is to improve the reusability of data for machines and software[14]. Several funders and governments worldwide advocate FAIR principles to maximize the value of collected data and to reduce research costs. For instance, the European Commission data management guidelines incorporated the notion of FAIR in 2017 [15]. According to a European Commission report, the negative consequence of not using FAIR research data is significant for the European economy [16]. However, the efforts to implement FAIR principles for management of health data optimized for machines remain in their early stage [15,17] and a framework to develop FAIR data specific to cystoscopic media contents is not well defined.

Given unmet needs, we present a conceptual framework for standardized management and quality control of cystoscopic images and videos, specifically in the setting of bladder cancer, a common cancer in which cystoscopy plays a central role in detection, treatment, and surveillance[18]. We further demonstrated the utility of this framework in real world scenarios for educational and research purposes in compliance with FAIR principles (FAIR data).

# Material and Methods

## The Framework Design

The conceptual framework is designed following FAIR (findable, accessible, interoperable, reusable) principles (**Table 1**) [14,19] and consists of three major sections as shown in **Figure 1**. The framework workflow is summarized in **Supplementary figure 1**. The study was approved by the Stanford University Institutional Review Board (Protocol #29838 and #36085) and written informed consent was obtained from all study subjects.

*Table 1 lists FAIR principles[14,19].*

| |
|---|
| **Findable** |
| F1. (Meta) data are assigned a globally unique and persistent identifier. |
| F2. Data are described with rich metadata (defined by R1 below). |
| F3. Metadata clearly and explicitly include the identifier of the data they describe. |
| F4. (Meta)data are registered or indexed in a searchable resource. |
| **Accessible** |
| A1. (Meta)data are retrievable by their identifier using a standardized communications protocol. |
| A2. The protocol is open, free, and universally implementable. |
| A3. The protocol allows for an authentication and authorization procedure, where necessary. |
| A4. Metadata are accessible, even when the data are no longer available. |

**Interoperable**

I1. (Meta)data use a formal, accessible, shared, and broadly applicable language for knowledge representation.

I2. (Meta)data use vocabularies that follow FAIR principles.

I3. (Meta)data include qualified references to other (meta)data.

**Reusable**

R1. (Meta)data are richly described with a plurality of accurate and relevant attributes.

R2. (Meta)data are released with a clear and accessible data usage license.

R3. (Meta)data are associated with detailed provenance.

R4. (Meta)data meet domain-relevant community standards.

## Data Management

The first section aims to manage the acquisition and storage of videos and images; the visual cystoscopy content is stored that the corresponding demographic and pathologic information are accessible while remaining compliant with patient privacy regulations. Each patient receives a unique identification number (UID) to facilitate data linkage between different data modalities and the pseudonymization procedure.

All videos are assigned to the corresponding patients based on their filenames as described in the supplementary section (standard protocols). All screenshots are labelled according to standard protocols that include additional context such as cystoscopy setting, bladder lesion location, and lesion pathology; these labels facilitate to find the visual contents in each patient. We considered vocabularies or abbreviations widely used by urological guidelines and community for pathology description and bladder anatomy [20-25], and partially in accordance with SNOMED terminology[26]. We did not apply the hierarchical system of SNOMED terminology or its coding system to maintain the human readability for these labels. Data management incorporates a quality control procedure to measure the initial data quality that will be described in “Swiss cheese model for Quality control” subsection. Image or video preprocessing (e.g., excluding blurred frames or images) is utilized to increase the content quality. A detailed information about documentation standard for is available in the supplementary section (standard protocols).

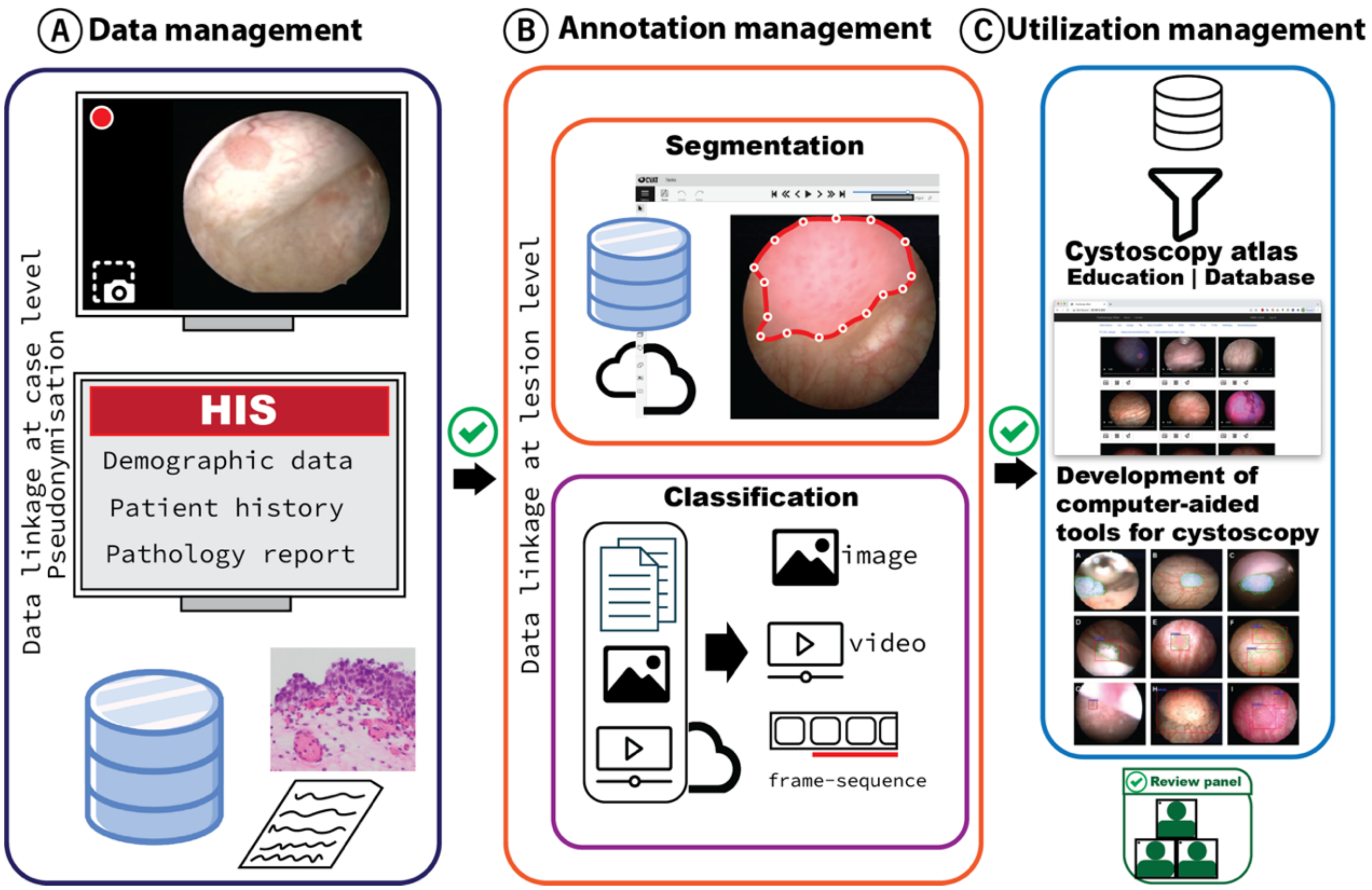

*Figure 1 The conceptual framework for the management of cystoscopic data. (A) data management for textual (clinical notes and pathology reports) and video data, (B) annotation management and (C) utilization management to facilitates resources for educational and research projects. The advancement from one section to another requires passing the quality control for each case. Annotation strategies are tailored by the goals of the utility management and can cover segmentation or classification problems. HIS: Hospital Information System.*

## Annotation Management

The goal of the second section is to manage the granularity-aware annotation process for lesions of interest found on cystoscopy and the resulting data. Lesions were labeled with information on the anatomical location of the lesion within the bladder (**standard protocol sections in supplementary material**), the appearance of the lesion (e.g., papillary, sessile, flat), and the pathological diagnosis of the lesion. The resources for annotation can be screenshots, single frames, short video clips or a full-length video of a cystoscopy procedure. Parallelly, the annotation procedures are determined by the machine learning problem definition. Accordingly, we defined two types of annotation:

1. Annotation for classification: annotation of entire representative frame or frame sequence of a single lesion with lesion information.
2. Annotation for segmentation and object detection: spatial annotation of individual lesions with the delineation of the lesion boundary in addition to the lesion information.

Data schemes for fine-granular annotations and tools should be designed to comply with the FAIR principles and the local patient privacy regulations. The annotation contents are defined with the instruction by a clinical supervisor. Further information about standard protocol for annotation can be obtained from the supplementary section. The supplementary material includes the handbook for lesion annotation.

### Utilization Management

The third section manages the utilization of the data from the previous sections. The utilization management includes functions to find, access, interoperate, and re-use data. In our case, we defined two courses of data utilization: the creation of a cystoscopy atlas for education and sharing annotated data to develop computer-aided tools for cystoscopy. For data utilization, a final pseudonymization step is executed to remove all patient-identifying information for data sharing. Parallelly, a final superficial quality check on random data points is done before releasing the data. The data utilization step also determines what content is displayed in the cystoscopy atlas or used in the development of machine-learning algorithms.

## Quality Control Model

We defined a Swiss cheese model for quality control to increase the likelihood of obtaining high-quality data in the end and to facilitate the conduction root cause analyses to mitigate potential repetitive errors in the annotation process. The first layer (QC1) assesses the quality of the video and images acquired for cystoscopy documentation, the alignment of textual and visual data with the outlines defined by our concept. The second layer (QC2) verifies data completeness. The last layer (QC3) randomly checks the annotation data for completeness. Each layer of quality control is verified again by the subsequent layer (e.g., at QC3, the contents of both QC1 and QC2 are reviewed again). Data that passes all three layers is deemed high-quality data that can be subsequently utilized for purposes such as the cystoscopy atlas or for research in machine-learning algorithms. **Supplementary Figure 2** provides the summary of the Swiss cheese model for quality control and the criteria to pass each quality control.

Frames are assessed for blurriness using the Laplacian operator[27] and for image quality using BRISQUE[28]. Frames showing cystoscopes outside the bladder are manually excluded using a commercial software (Adobe Premier Pro; Adobe, San Jose, U.S.A). By fine-granular annotation of full-length videos, frames overlooked during the manual processing of videos are marked as excluded during the annotation process. Frames representing the transurethral resection of bladder tumors were marked as TURBT frames. The video quality was measured using a reference-free video assessment introduced by Li et al [29].

Following the Delphi method [30], annotation quality control was performed by a review panel consisted of clinical research coordinators, data scientists, urologists, and pathologists for content, label, and annotation qualities. Generally, all annotations should receive the approval from the majority (in our case: 3/4 urologists). In case of conflict in annotation, the team leader (i.e., principal investigator) is the resolving vote. In cases where external expertise is required to solve a conflict (e.g., to clarify pathology of a case), the team leader finds and consults the appropriate expert (e.g., reference pathologist) inside the primary institution to resolve the conflict.

## Framework Infrastructure

### Serverless solutions

The serverless solutions delegate the sever maintenance to third parties. We considered commercial solutions provided by our academic institution as infrastructure for data management (**Figure 2**). For example, we utilized BOX (BOX, Redwood City,

U.S.A.) that provides solutions for HIPAA-compliant data storage and management. Furthermore, this solution facilitates the access and search to the contents using the build-in File Explorer of the widely used operating systems (i.e., Windows and macOS) or the search tool from the vendor. Moreover, the serverless file storage facilitates running scripts to index the files. **Table 2** lists the serverless solutions we utilized for different data modalities. We emphasize that the access to these solutions is regulated by the institutional account infrastructure (i.e., Stanford University and VA). Finally, these serverless solutions provide API (Application Programming Interface) for interoperability; we applied a serverless file storage solution to sustain the availability and maintenance of our customized R and python scripts for the whole data management team.

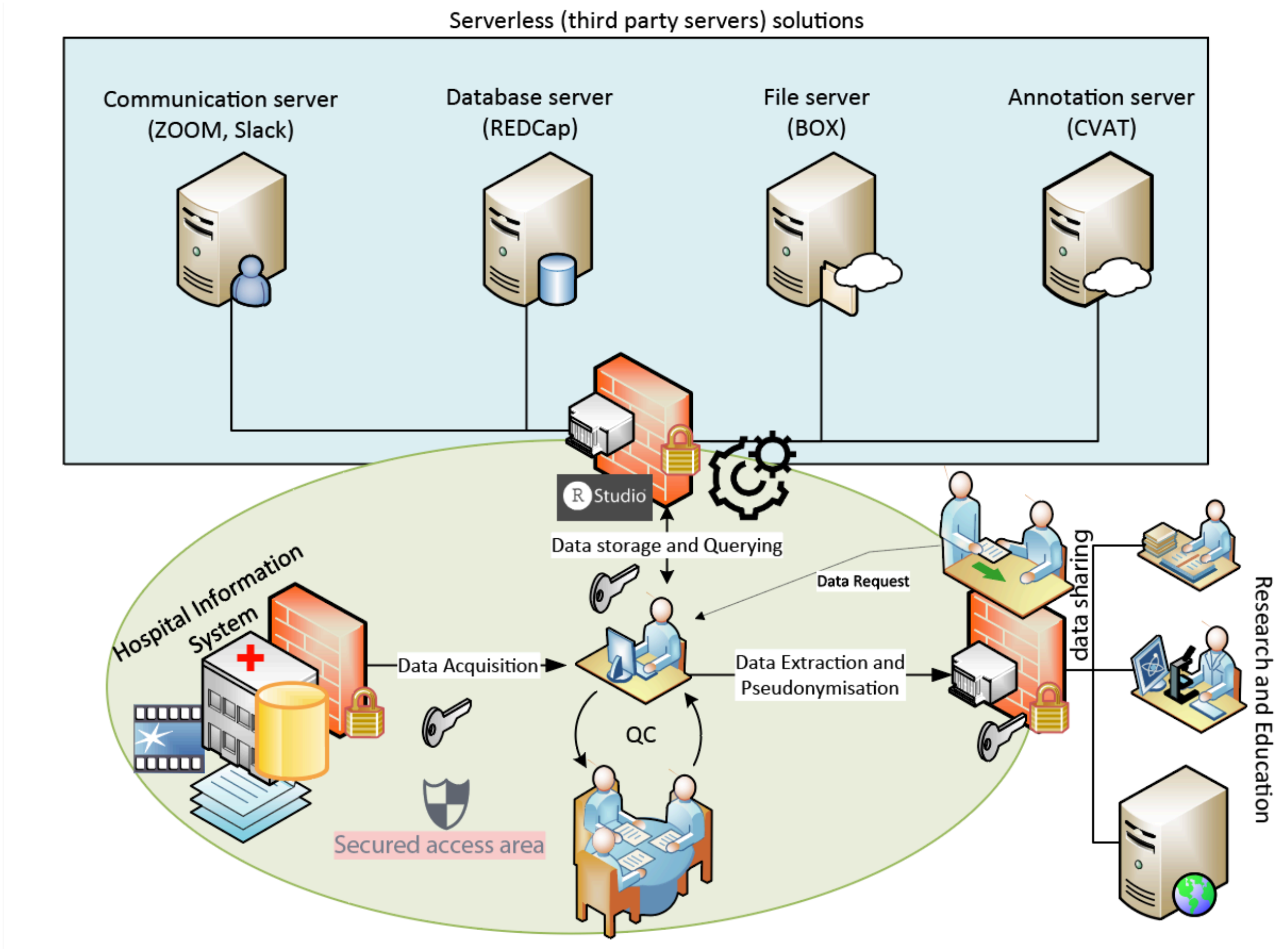

*Figure 2 summarizes the framework infrastructure and its information flow model. Three components of the infrastructure are highlighted as following: blue rectangle represents the serverless solutions, the green ellipsoid represent the information security policy of the infrastructure; the flow diagram represent the data flow model. We utilized serverless solutions for data storage to delegate the hardware maintenance and applied available communication platforms to coordinate the team efforts for data acquisition, annotation tasks or quality control. The secured access area is restricted to those who have the access permission and the two-factor institutional authentication. The secured access area includes data acquisition, data query and quality control, extraction and pseudonymization. A third party (research or education) can contact the principal investigator to request shared data*

*through email. Within the secured access area, we defined six units to execute the framework concept that are illustrated in Figure 3. The red walls symbolize the requirement for access permission. Home-customized scripts were used to manage multi-modal data sources.*

*Table 2 lists the serverless solutions used for the framework infrastructure.*

| Task | Serverless solution | Headquarters' location |
|---|---|---|
| Communication<br>Video conference<br>Threads | <br>ZOOM<br>Slack | <br>San Jose, USA<br>San Francisco, USA |
| Database storage | REDCap | Nashville, USA |
| File storage | BOX | Redwood City, USA |
| Annotation | CVAT -Computer Vision Annotation Tool-,<br>(OpenVINO, INTEL) | Santa Clara, USA |

## Information Security Policy

In accordance with the patient's privacy regulations[31,32], we defined a secured access area in the framework infrastructure and policies to prevent unauthorized access to sensitive data. Four data levels were outlined according to ISO 27001 information classification[33] (**Supplementary table 4**). All users should have an appropriate permission level to gain access to the data source and are controlled by institutional infrastructure. Furthermore, we limited this first and second level permissions (restricted and confidential data) to the team members whose tasks are to admin the multimodal databases. A link to de-identified data is provided to internal and external users after obtaining approval by the principal investigator (JCL). The shared data should meet the fourth data level (public data) for external users and third data level (interior data) for internal (e.g., members within the department for urology at Stanford or VA) users.

## Information Flow Model

The information flow model describes the data flow within the framework and is secured according to the security policy described earlier (**Figure 2**). Textual data (e.g., demographic information, smoke exposure, pathology and surgery reports) were extracted manually from the electronic health records. Parallelly, video records of cystoscopy and screenshots of lesions were obtained during the cystoscopic procedures. The electronic health records and raw cystoscopic visual contents are only accessible to members listed in the IRB approval.

We implemented a customized guided user-interface (GUI) tool to standardize the entry of cystoscopic visual contents on the cloud file storage (**Figure 3**). The user-interface tool for images includes autocomplete input fields for study identification, lesion pathology, lesion location and image modality. The video files are automatically named based on the study identification and the acquisition date when transferred to the cloud file storage.

To trace the label completion for images, we defined a list of status reflecting the data completion (**Supplementary Table 1 and 2**). A review tool was also developed to browse the images according to the data completion status with editing functions. Furthermore, the data entry clerk can access to the case folder with the corresponding videos and images to add new visual content if needed. Parallelly, the file storage documents the changes in these folders and facilitates file recovery for a period of 180 days. The workflow for quality control is followed as described earlier in the framework concept. We incorporated customized scripts that index the image files in the cloud storage to facilitate the content search and cohort curation based on these images. The list of indexed files is stored in csv format (comma-separated values) and is interoperable with other script languages like R and python. As part of HIPAA- compliant data sharing, we provide a de-identified version of the file list for external and internal data requesters. The infrastructure to implement the information flow model is explained in the supplementary material.

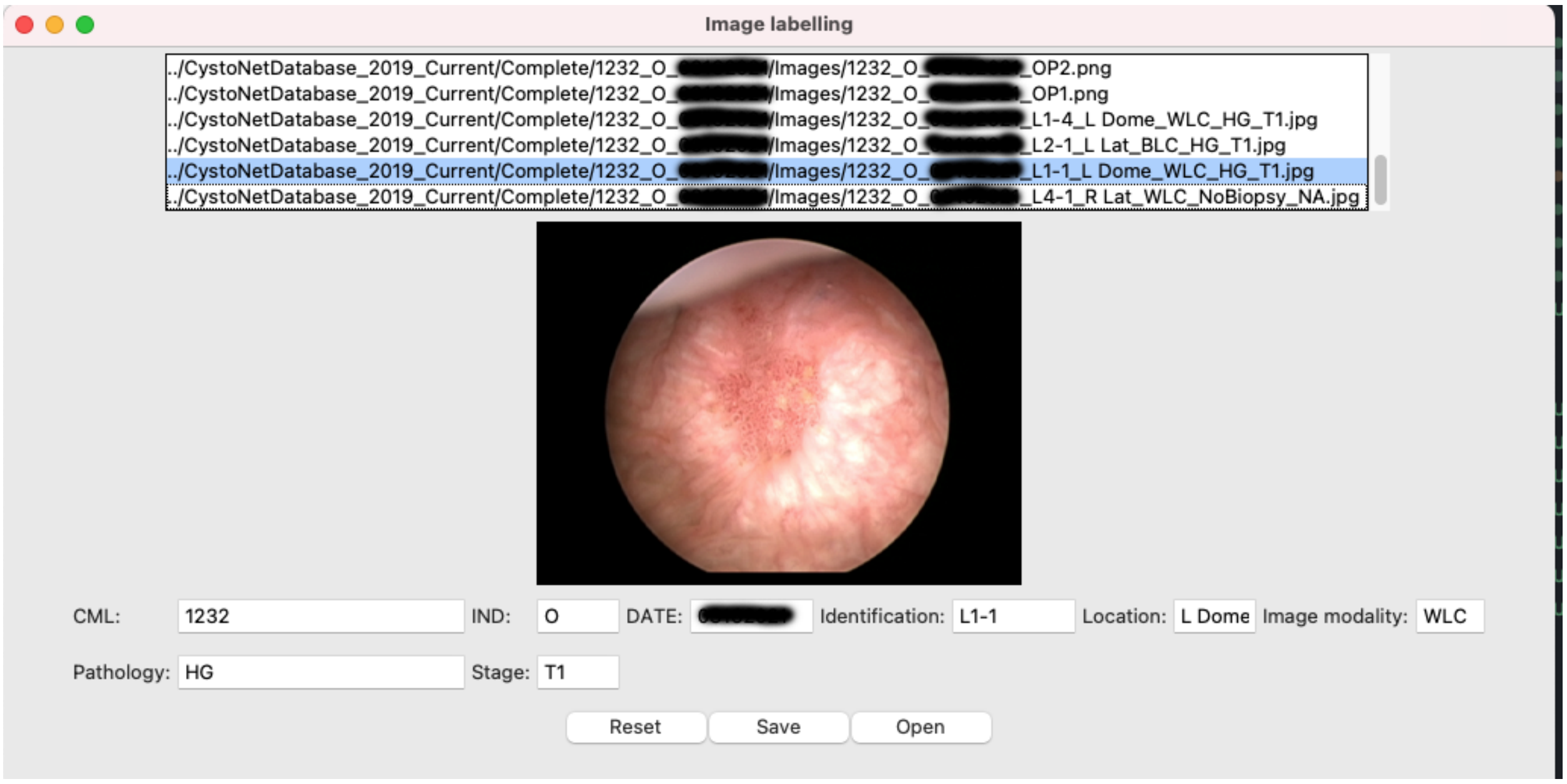

*Figure 3: The user guided interface for image labeling with autocomplete input fields. Supplementary table 1 and 2 provide possible input options for each input field. The GUI tool is a python script that is compatible with most operating systems.*

## Evaluation

The success of the framework implementation was measured by following FAIR (Findable, Accessible, Interoperable, Reusable) principles. To demonstrate the operability and implementation feasibility of the framework, we examined the framework by tracing the curation of two distinct data sets to mimic two scenarios:

1. Data sharing for research projects
2. Educational web portal

The cohort description of full-length cystoscopy videos for data sharing included the lesions and the video frames annotated for these lesions.

## Results

The conceptual framework was successfully implemented utilizing infrastructure from two different sites (Stanford Hospital and Palo Alto VA). The implementation procedure required approximately two and half months until having a fully operational framework that mostly involved the allocation of the personal resource and building and checking the computational infrastructure for fault tolerance.

The video and image contents for 123 TURBT cases, existing prior the framework implementation, were retrospectively processed and then managed by the framework. In addition to that, we prospectively added 156 TURBT cases and 91 clinical cystoscopy cases in the period between Jan. 2019 and May 2022. The media contents (videos and images) of 370 cases required 449.44 GB for storage.

## Framework evaluation

### Findable

The framework facilitated the key search in 371 images with complete labels from 65 cases who were enrolled in the period between Jan. 2020 and May 2022; we could find images representing different pathology using the metadata generated within the framework (**Figure 4**). **Figure 5** illustrates some example images representing different lesion pathologies and image modalities.

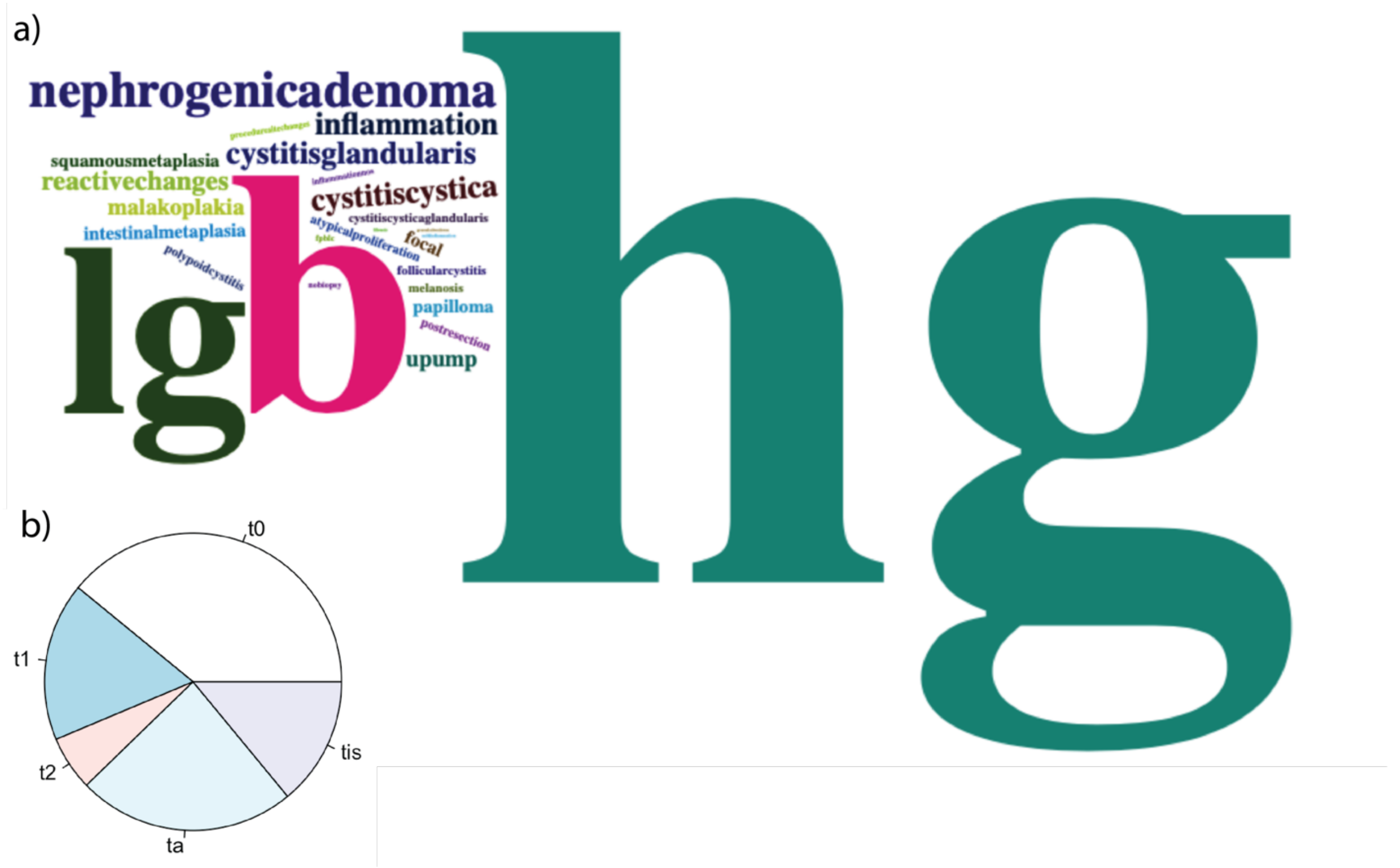

*Figure 4: a) the searchable keys and their frequencies (boldness) are presented as word clouds. Hg: high-grade bladder cancers, lg: low-grade bladder cancers, b: benign, unspecified. UPUMP: Urothelial proliferation of unknown malignant potential. b) the distribution of pathologic tumor stages in 371 images.*

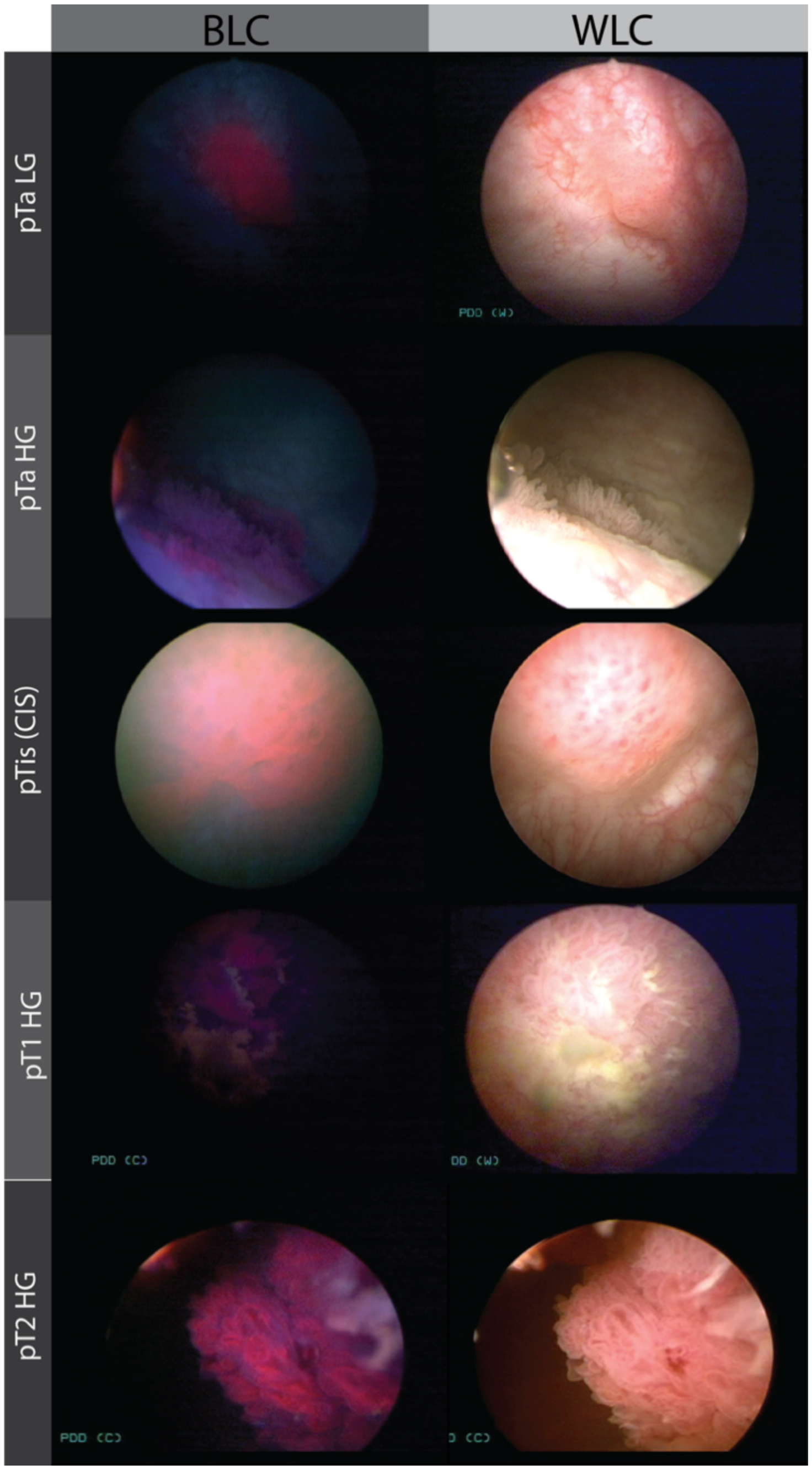

*Figure 5 illustrates bladder cancer lesions visualized using blue light (BLC) and white light (WLC) image modalities.*

*Here, we used the keywords described in Supplementary table 3.*

## Accessible

All images, videos and corresponding metainformation were accessible on the cloud file storage using the framework for authorized users. Metainformation includes storage information about the image and video and therefore facilitates authorized users to access to the visual contents. A direct access to data without having the appropriate permission level is not feasible.

## Interoperable

Given that the segmentation annotation of lesions is time- and labor-intensive, we selected videos from 68 TURBT cases with 163 pathology-confirmed bladder cancer or benign lesions from the prospectively enrolled case cohort (n=156), achieving an approximately uniform distribution of benign lesions, low-grade bladder cancer, and high-grade bladder cancer (**Table 3**). We uploaded these full-length cystoscopy videos with pathology-proven lesions on CVAT for segmentation annotation; here, we wrote a python script that watches for new video files in a folder and upload them on CVAT using CVAT API in our internal network. After completing the annotation task, these annotations were accessible using API and can be extracted for lesion-level analyses. Documentation standards meet the domain relevant standards, unify the content communication between different software and tools (i.e., R studio); the framework facilitated the data integration at different levels (i.e., frame, lesion and case) as shown in **Table 3** and **4**.

*Table 3: The cohort description for segmentation annotation. IQR: interquartile range.*

| | Description |
|---|---|
| Patients, n | 60 |
| Cases, n | 68 |
| Lesions, n | 163 |
| Lesions per case, median (IQR) | 2 (1 – 3) |
| Pathology, lesion n (%) | |
| Benign | 49 (30.1) |
| Bladder Cancer | 114 (69.9) |
| Tumor stage, n (%) | |
| Ta | 70 (61.4) |
| CIS | 23 (20.2) |
| T1 | 13 (11.4) |
| T2 | 8 (7.0) |
| Tumor grade, n (%) | |
| Low grade | 54 (47.4) |
| High grade | 60 (52.6) |

*Table 4: The frame distribution by lesion pathology for 68 cystoscopy videos.*

| Pathology | Frames, n (%) |
|---|---|
| Background | 263,897 (74.6) |
| Benign | 28,954 (8.2) |
| Bladder Cancer | 60,830 (17.2) |
| Tumor stage | |
| Ta | 28,472 (46.8) |
| CIS | 14,457 (23.8) |
| T1 | 14,348 (23.6) |
| T2 | 3,553 (5.8) |
| Tumor grade | |
| Low grade | 18,420 (30.3) |
| High grade | 42,410 (69.7) |

## Reusable

### *Research*

The previous cases with full-length videos underwent transurethral resection of bladder tumors in the period between 2019 and 2021. The videos are edited prior the annotation to exclude frames not related to the endoscopic procedures. Frames suffered from poor image quality due to the surgery conditions were also excluded. Overall, these videos included 857,032 frames and their resolution ranged from 320p to 1080p.

We annotated a total of 353,681 (41.3%) frames covering cystoscopy examination after excluding frames representing the resection procedures. The annotated data with polygon shapes and lesion pathology are extracted in COCO format [34] for each video. Both videos and corresponding COCO formats were provided for a research project focused on developing computer-aided endoscopic solutions. The metainformation file reveals the curation and modification date. An example file for COCO file is provided in the **Supplementary file 1**. **Table 5** lists the research projects utilizing our FAIR data; The results of these project were peer-reviewed and accepted for presentation in SPIE Photonics West (San Francisco, U.S.). The conference paper for each project will be available in 2023.

*Table 5 lists the three projects utilized our FAIR data and their conference papers were accepted after a peer-review process.*

| Project name [Reference] | Project summary | Purpose (s) | annotation type |
|---|---|---|---|
| Sequential modeling for bladder tumor classification [35] | training and validating four state-of-the-art sequential models to differentiate between bladder regions that merit biopsy and normal urothelium. | -Model development<br>- Validation | Classification |
| Flat lesion detection of white light cystoscopy with deep learning [36] | The project applies deep learning algorithm for augmented flat lesion detection of WLC. The algorithm was designed by incorporating domain adaptation, transfer learning, and region of interest detection. Without relying on BLC, our algorithm can produce flat lesion predictions very close to urologist's annotations. The proposed deep learning algorithm holds strong potential to improve performance of WLC in a noninvasive and cost-effective fashion. | -Model development<br>- Validation | segmentation/ object objection |
| Potential of educational cystoscopy atlas for augmented intelligence [37] | The recognition of cystoscopic findings requires a prolonged learning curve and depends on the examiner's skills. Although computer-aided detection tools hold the | -Validation | classification |
| | potential to improve the performer's experience, compiling a comprehensive dataset for such tools remains a challenge. Educational cystoscopy atlas represents an alternative strategy to overcome this challenge. The project utilized an educational cystoscopy and a neural architecture search to develop deep learning models for bladder cancer detection with reasonable accuracy. The preliminary results show the potential utility of educational cystoscopy atlas for augmented intelligence. | | |

### *Education*

For the cystoscopy atlas, we considered 371 still images from complete cases who were enrolled in the period between January 2019 and Mai 2022. The metainformation file included lesion identification, pathology, and location for each still image; the curation and modification date can be obtained from the metadata of the metainformation file. **Figure 6** provides a screenshot of the web portal for cystoscopy atlas at case level for internal users. **Supplementary file 2** presents an example for metainformation file used for curation of the cystoscopic atlas.

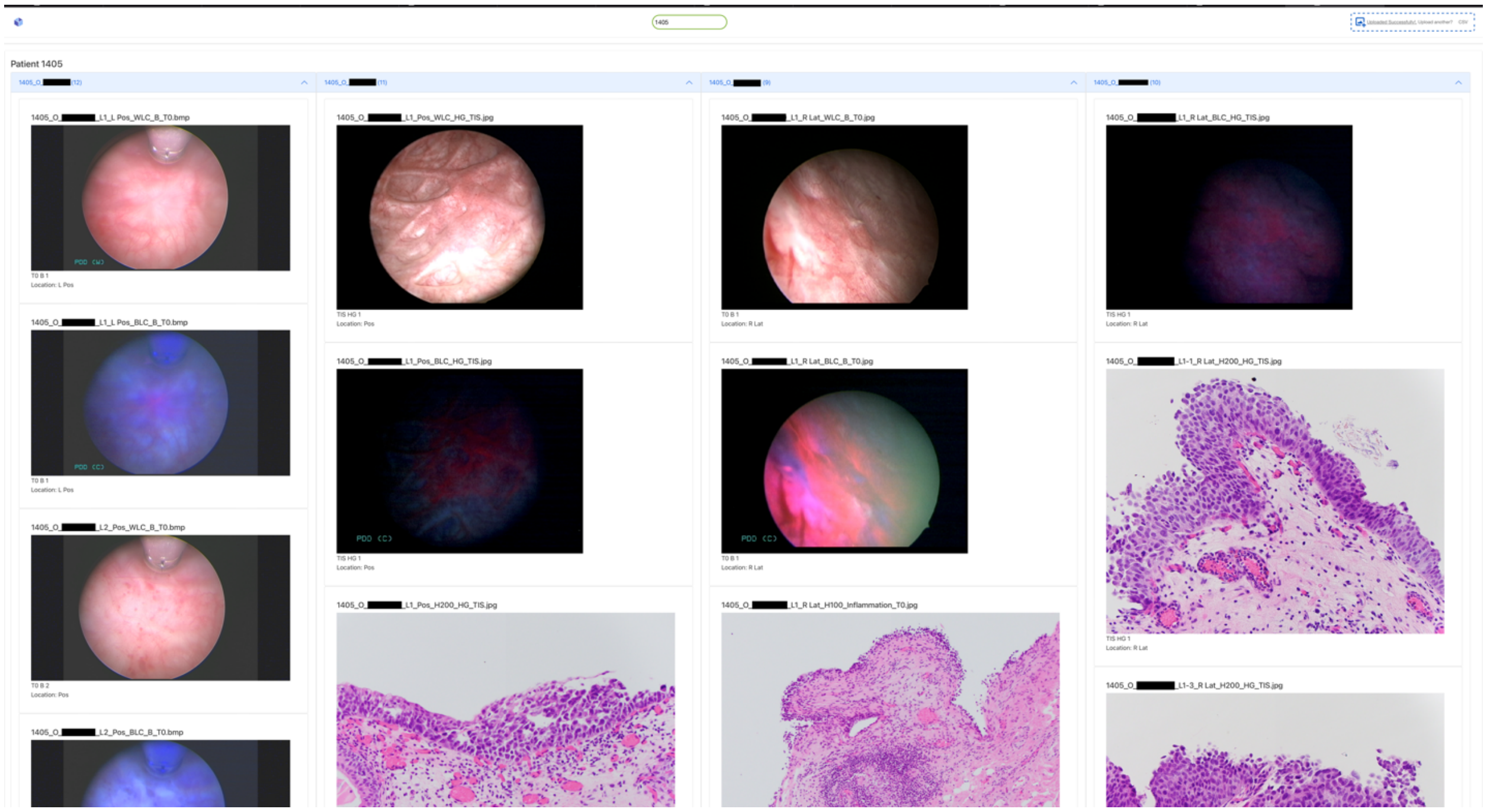

*Figure 6 illustrates the internal web portal for cystoscopic atlas. Here, the cases belonging to a single patient are presented sorted by recent date (left most recent, right: oldest). BLC: the blue light cystoscopy, WLC: white light cystoscopy. The web portal includes options to search (upper center) or upload comma separated file (upper right) to view selective images. The textual pathology and surgery reports are presented for each case as well.*

## Quality control

We reviewed 68 full-length cystoscopy videos and the corresponding annotation. Overall, the median BRISQUE score for frame quality is 57.1 (IQR: 37.7 – 78.4) and the median video quality score was 85.9 (IQR:  79.8– 90.9).

QC2 identified and corrected the annotation and labels in 12 lesions (7.4%). **Table 6** list causes for errors in the annotation task according to the root cause analyses. Here, the major issues were the difficulty in identifying the lesions described in the surgery report or the screenshots, the high-grade multifocality of the tumor that challenges the annotators and small lesions that does not provide enough tissue samples for pathology evaluation. After QC2, data completeness and the annotation agreement for these lesions had reached %100.

*Table 6 lists results of the root cause analyses conducted while performing the annotation tasks.*

| Causes for not including lesions for annotation |
|---|
| Difficulty in delineating all multifocal lesions in a single case with >5 lesions. |
| Lack of pathology information due to tissue sampling |
| Difficulty in identifying the lesions by reviewing a full-length (~>10-30 minutes) video |
| Poor quality of the frames |
| **Causes for incorrect lesion labeling** |
| Failure in associating the pathology to the lesion |
| Lack of pathology information due to tissue sampling |
| Lack of pathology information in REDCap due to data incompleteness (Rechecking) |
| **Causes for incorrect lesion delineation** |
| Rapid motion of the camera view field |
| Difficulty in identifying the flat lesion boundary |
| Large lesion appearance |
| The camera was too close to the lesion |
| The camera was too far from the lesion |
| Poor frame quality |

## Discussion

The current study successfully demonstrates a framework that acquires, maintains, and consumes cystoscopic visual data following the FAIR principles. Our framework proposes documentation standards that facilitate data integrity and the secondary use of cystoscopy's visual contents. Furthermore, the current work advocates strategies to measure the data quality of cystoscopy-related contents as data quality plays an essential role for computer vision and machine learning [38-41]. Collecting high quality data requires a well-thought-out and assessed conceptual framework with rigorous quality controls. Furthermore, the framework proposes standards to manage the cystoscopic visual documents in a human readable format.

Although information is the central component of artificial intelligence research, data engineering is the most underestimated factor in this research domain. Data quality is rarely reported in the literature for medical artificial intelligence possibly due to its unknown effect on model performance and weighing model generalization over the data quality [42,43]. For instance, previous efforts on AI-augmented bladder tumor detection primarily focused on the development of deep learning models with high accuracy while softening the principle of data quality and data generalization[6,7,44-47]. Another possible assumption for underestimating data quality is considering data quantity in order to compensate model generalization issues by applying more data. However, increasing data quantity or having large datasets is not necessarily associated with a better model performance nor model generalization [48-50].

In the last decades, data management received special attentions from the industry due to its relevance for stakeholders and consumers to improve the user-experience and the production, such that a global standard for data quality and enterprise master data (ISO-8000) was introduced [51]. However, this standard is tailored to business and provides an abstract concept for data exchange and role assignments that require extensive human and infrastructure resources, therefore not optimized for clinical facilities and research-based projects with limited budgets. In contrast, our work enables the implementation of the conceptual framework upon existing infrastructure to deliver a qualitative cystoscopy atlas. Notably, this cystoscopy atlas has reflected a wide range of cystoscopic findings. In addition to the effort on defining a good cystoscopy by urologic guidelines [52], our framework is applicable to optimize the data acquisition and clinical documentation of endoscopic imaging; the proposed framework is expansible and scalable to manage data from multi-sites as well.

Data acquisition during clinical setting is generally suboptimal and requires appropriate quality control and standards to elevate data quality. However, the metrics and measurements for qualitative data should be realistic and reflective of data quality for clinical routine to avoid extreme levels of data quality not representative of real-world conditions.

One of the major goals of FAIR principles is the automation of artificial intelligence solutions by providing data readily available for direct use and scaling. This is essential for the life cycle development of clinical decision-aided tools. We therefore considered FAIR principles, while establishing the framework, to build a foundation for artificial intelligence (AI) platforms that automate diagnosis procedures and repetitive documentation tasks in urologic endoscopy. We further showed the real-world utility of our framework in research as we could conduct three research projects related to AI for urologic endoscopy using our FAIR data.

We considered the cloud-based solutions due to its efficient scalability and the delegation of storage maintenance to a third party. Downscaling to a local network solution is also feasible. Our python scripts to manage data entry and the R project files are stored where the image files are stored and therefore can be executed by remote clients according to their user permission level.

Our proposed framework requires the file explorer of the operating system and the cloud file storage drive that facilitates the access to the cloud files. Our framework is compatible with any HIPAA-compliant cloud solution that facilitates mounting the cloud fie storage as a drive in an operating system and parallelly regulates the cloud file access.

Instead of the cloud file storage drive, the framework can utilize a local network drive to store files according to standards defined by the framework. However, the network owner should be actively involved in achieving HIPAA compliance at user, software and hardware levels.

The open-source annotation tool CVAT runs on a cloud-based virtual machine. Accordingly, we can intuitively replace this virtual server with a local server to operate the annotation tool that is accessible within the local network. We emphasize that the annotation tool should provide API to achieve an efficient data transfer  or exchange.

We do not expect any issues with considering alternative products for communication and database management that are HIPAA-compliant. Moreover, any HIPAA-compliant database management can be used when the database scheme is designed in accordance with the framework concept. Finally, we preferred the open-source statistical and data mining tool “R studio” as two versions of this tool are available for desktop and server and well-known for data science community.

Although our framework is effective in generating high-quality data, there are inherent limitations. The implementation of the conceptual framework requires high-skilled engineers and technical resources at the institution. The infrastructure maintenance or operation is associated with cost and human resources. However, the benefits of having a framework for high-quality data overweigh its harms, especially in residency education [53] and artificial intelligence research that requires high-quality and representative data.

In summary, the current work introduces a framework that facilitates the curation of qualitative data for endoscopic artificial intelligence research. Future works will focus on integrating artificial intelligence solutions to automate repetitive tasks of the frameworks.

## Conclusion

We introduced a novel conceptual framework that provides standards to store and manage the media documentation of cystoscopy procedure. Using this framework further facilitates the development of high-quality cystoscopy atlas as secondary use for research and education.

## Acknowledgement

The research was supported in part by National Institutes of Health R01 CA260426 to JCL.

## Data availability

Due to the institutional policy, data utilized in this work cannot be publicly shared. However, a query can be submitted to the corresponding author JCL for external data sharing.

## Code availability

The python scripts for image labeling and quality control can be obtained from https://github.com/oeminaga/framework_cystoAI_fair. CVAT can be obtained from https://github.com/openvinotoolkit/cvat. The script for video quality score can be downloaded from https://github.com/lidq92/VSFA.

## Author contributions

Okyaz Eminaga: Conceptualization, Methodology, Software, Investigation, Data curation, Visualization, Writing- Original draft preparation.

Timothy Jiyong Lee: Data curation, Investigation, Visualization.

Jessie Ge: Data curation, Investigation.

Eugene Shkolyar: Writing- Reviewing and Editing.

Mark Laurie: Software, Data curation and Visualization.

Jin Long: Validation.

Lukas Graham Hockman: Data curation.

Joseph C. Liao: Data curation, Investigation, Supervision.